\documentclass[10pt]{article}
\usepackage{amssymb,verbatim,amsmath}
\numberwithin{equation}{section}

\newtheorem{theorem}{Theorem}[section]

\newtheorem{remark}[theorem]{Remark}
\newtheorem{corollary}[theorem]{Corollary}

\newtheorem{definition}[theorem]{Definition}

\newtheorem{ex}{Example}[section]

\newtheorem{ass}{Assumption}[section]

\numberwithin{equation}{section}

\begin{document}

\newcommand{\ddt}{\partial \over \partial t}
\newcommand{\ddx}{\partial \over \partial x}
\newcommand{\ddu}{\partial \over \partial u}

\newcommand{\ddtbar}{\partial \over \partial \bar{t}}
\newcommand{\ddxbar}{\partial \over \partial \bar{x}}

\begin{center}
{\Large {\bf Symmetries of Kolmogorov backward  equation}}
\end{center}

\bigskip

\begin{center}

{\large Roman Kozlov}

\medskip
\hspace{1.5 cm}

Department of Finance and Management Science, \\
Norwegian School of Economics, \\
Helleveien 30, 5045, Bergen, Norway; \\
E-mail address: Roman.Kozlov@nhh.no \\

\end{center}


\bigskip

\bigskip
\begin{center}
{\bf Abstract}
\end{center}
\begin{quotation}
The note provides the relation between symmetries and first integrals of 
It\^{o}  stochastic differential equations 
and symmetries of the associated Kolmogorov backward  equation. 
Relation between the symmetries of the Kolmogorov backward  equation 
and the symmetries of the Kolmogorov forward equation  is also given. 
\end{quotation}

\bigskip

Key words:

\bigskip

Lie symmetry analysis

Stochastic differential equations

Kolmogorov backward equation










\bigskip
\bigskip




\section{Introduction}
\label{Intro}

Lie group theory of differential equations is well
developed~\cite{Ovs, Ibr1, Olver}.  
It studies transformations which take solutions of differential equations into other solutions of the same equations. 
This theory became a powerful  tool for finding analytical solutions of differential equations.

Successful applications of  Lie group theory to differential equations motivated 
its development for other equation types. 
Here we consider stochastic differential equations (SDEs). 
First attempts were devoted to transformations which change only the dependent variables, 
i.e.  transformations which do not change time~\cite{Mis1, Mis2, Alb}. 
After them fiber-preserving transformations  were approached~\cite{Gae1}.    
Later there were considered general point transformations  
in the space of the independent and dependent variables~\cite{Wafo, Unal1, Fre, Shi, Fre2}. 
For them the transformation of the Brownian motion is induced by the random time change.   
We refer to a review paper~\cite{Gaetareview}  for symmetry development 
and to a chapter~\cite{Kozlov9} for symmetry applications.  
More general framework includes transformations 
which depend on the   Brownian motion~\cite{Gaeta2000, Spadaro,  Kozlov7, Kozlov8}.

One of the applications of the symmetries of SDEs was their  relation to 
symmetries of   
the associated  Kolmogorov forward equation (KFE), 
which is also known as the  Fokker--Planck equation in physics~\cite{Risken}. 
First,  this symmetry relation was treated in~\cite{Gae1} for fiber-preserving symmetries. 
Later,  it was considered for symmetries in the space of the independent and dependent variables in~\cite{Unal1}.
In~\cite{Kozlov5}  (see also~\cite{Kozlov6}) a more precise formulation of the symmetry relation  was provided.
There is also a relation between first integrals of the SDEs and symmetries of the KFE~\cite{Kozlov5}.   
There are many papers devoted to symmetries of particular Fokker--Planck equations  
 \cite{Sastri,  Sht,   Cic1, Cic2, Rud,    Fin,  Spichak}.

Symmetries of the Kolmogorov backward  equation (KBE) received much less attention than symmetries of the KFE.  
The Kolmogorov backward  equation is useful when one is interested whether 
at some  future time the system will be in a target set, i.e.  in a specified subset of states.    
In~\cite{Ugolini_1} the authors considered symmetries  of a KBE with  diffusion matrix 
(matrix $A_{ij}$ in Eq.~(\ref{BK})) of full rank. For such equations corresponding to autonomous SDEs 
and time changes restricted to scalings  it was shown that symmetries of the SDEs are also symmetries of the KBE.   
The paper~\cite{Ugolini_2} examines more general stochastic transformations  able to change 
the  underlying probability measure. In this framework the weak extended symmetries of SDEs 
are more general than the Lie point symmetries of the Kolmogorov backward  equation.

In the present note we consider Lie point symmetries of the Kolmogorov backward  equation 
and examine how these symmetries can be related to the strong symmetries of the underlying SDEs 
without the restrictions which were imposed in~\cite{Ugolini_1}. 
We also consider the Lie point symmetries of the KBE 
corresponding to first integrals of the  underlying SDEs   
and show how  the symmetries of KBEs  are related to  the symmetries of KFEs 
corresponding to the same underlying SDEs.

The paper is organized as follows. 
In the next section we recall basic results on It\^{o}  SDEs and their symmetries. 
In section~\ref{section_backward}   we examine Lie point  symmetries 
of the Kolmogorov backward  equation 
and find out how they can be related 
to the symmetries of  the stochastic differential equations 
and to the symmetries of the Kolmogorov forward equation. 
Finally, in section~\ref{section_scalar} we consider scalar SDEs and $(1+1)$-dimensional  Kolmogorov equations 
to illustrate the theoretical results of this paper.  
The last section also illustrates the theory on an example of geometric Brownian motion.


\section{SDEs and Lie point symmetries}

\label{main_theory}

Let us consider a system of  stochastic differential equations in   It\^{o} form 
\begin{equation}    \label{sys} 
dx_i = f_i ( t, {\bf x}) dt + g_{i  \alpha } ( t, {\bf x}) dW_{\alpha} (t)  , 
\qquad 
i = 1, ..., n , 
\qquad 
\alpha = 1, ..., m    , 
\end{equation}
where $ f_i (t,  {\bf x} ) $ is a drift vector, 
$ g_{ i \alpha} (t, {\bf  x} ) $ is a diffusion matrix 
and $ W _{\alpha} (t) $  is a vector Wiener process (vector Brownian motion)~\cite{Arn, Evans, Gard,  Ikeda, Oks}. 
We assume summation over repeated indexes 
and use notation $ {\bf  x}  = ( x_1, ..., x_n ) $. 
Let us remark that $ W_{\alpha} (t)  $, $ \alpha =1, ..., m $   are independent one-dimensional Brownian motions.

\subsection{It\^{o} formula}     

\label{formula}

The transformation of the dependent variables in stochastic calculus is given by It\^{o} formula  
(see,  for example,~\cite{Oks}). 
For SDEs~(\ref{sys})   we perform variable change 
$  {\bf x }  \rightarrow    {\bf y }  = {\bf y }  ( t, {\bf x}) $  
according to 
\begin{equation}     
d y _ i   
=    { \partial  y _i   \over \partial  t}   dt   
+   { \partial  y _i   \over \partial  x_j }  d x_j   
+  { 1 \over 2 }     { \partial  ^2 y _i   \over \partial  x_j   \partial  x_k }    d x_j  d x_k    , 
\qquad
i = 1, ..., n   ,
\end{equation}     
where $  d x_j  d x_k $ are found with the help of the substitution rules 
\begin{subequations}   \label{rules} 
 \begin{gather}
dt \cdot dt = 0 ,      \label{rule1}   \\
dt  \cdot  d W _{\alpha}  =    d W _{\alpha}  \cdot  dt  =  0 ,    \label{rule2}   \\
d W _{\alpha}  \cdot   d W _{\beta }  = \delta _{ \alpha \beta }  dt      .     \label{rule3} 
 \end{gather}
\end{subequations}
Thus, we obtain the formula  for differentials  in stochastic calculus 
\begin{equation}     \label{differential} 
d F (t,  {\bf x} )  = D_0 ( F )  dt  + D_{\alpha}  ( F)  dW_{\alpha} (t)     , 
\end{equation}
where 
\begin{equation}  \label{intcond1_ito}
D_0   =  { \partial   \over \partial  t}
+ f_j { \partial   \over \partial  x_j }
+ { 1 \over 2 } g _{j \alpha } g _{k \alpha }  { \partial ^2     \over \partial  x_j \partial  x_k }  , 
\qquad
 D_{\alpha}   = g _{j \alpha } { \partial    \over \partial  x_j }   . 
\end{equation}

\subsection{First integrals}     

\label{integrals}

Stochastic differential equations can possess first integrals.

\begin{definition}    \label{def_first_integral}
A quantity $ I ( t,  {\bf x}) $ is a first integral of a system of SDEs~(\ref{sys})
if it remains constant on the solutions of the SDEs.
\end{definition}

Application of the It\^{o} differential formula~(\ref{differential}) to a first integral
\begin{equation*}
dI (t,  {\bf x})  = D_0 (I) dt  + D_{ \alpha }  (I)   dW_{\alpha} (t)   = 0
\end{equation*}
leads to a system of partial differential equations
\begin{subequations}   \label{intcond1_I}
\begin{gather}
 D_0 (I)  = 0 ,     \label{intcond1_I_a}   
\\
 D_{ \alpha }  (I)  =  0    .    \label{intcond1_I_b}
\end{gather}
\end{subequations}




\subsection{Determining equations}

\label{Determining}

We will be interested in infinitesimal group transformations
(near identity changes of variables) in the space of the independent and dependent variables 
\begin{equation}      \label{transformations}   
\bar{t} = \bar{t} (t,{\bf x},a) \approx t  +  \tau (t,{\bf x})  a,
\qquad
\bar{x}_i  = \bar{x} _i (t,{\bf x},a) \approx x_i   +  \xi _i  (t,{\bf x})  a,
\end{equation}  
which leave Eqs.~(\ref{sys})
and framework of It\^{o} calculus invariant.
Such transformations can be represented by
generating operators of the form
\begin{equation}   \label{symmetry}
X = \tau (t,{\bf x}) {\ddt} + \xi _i ( t,{\bf x} ) {\partial \over \partial {x_i}} .
\end{equation}

The determining equations
for Lie point transformations~(\ref{transformations})   of It\^{o} SDEs~(\ref{sys}) 
were derived in~\cite{Unal1}.  
It is convenient to present them with the help of the operators   $D_0$   and $ D_{\alpha}$ 
given in~(\ref{intcond1_ito}). 
The determining equations take a compact form
\begin{subequations}  \label{cond0new}
\begin{gather}
D_0 ( \xi_i ) - X ( f_i ) - f_i D_0 ( \tau ) = 0 ,     \label{cond1new} 
\\
D_{\alpha} ( \xi_i ) - X ( g_{ i \alpha} ) - {1 \over 2}  g_{ i \alpha} D_0 ( \tau ) = 0 ,    \label{cond2new} 
\\
D_{\alpha} ( \tau ) = 0 .   \label{cond3new} 
\end{gather}
\end{subequations}

In~\cite{Kozlov7}  it was shown that one can also obtain these determining equations 
by restriction of more general transformations 
which involve Brownian motion.    
The Lie point symmetries~(\ref{symmetry})  of It\^{o} SDEs, 
which are given by the determining equations~(\ref{cond0new}), 
form a Lie algebra~\cite{Kozlov5}.


\section{Symmetries of Kolmogorov  backward equation}

\label{section_backward}

In this section we derive the determining equations 
for Lie point symmetries of the Kolmogorov  backward equation  
and find out how these symmetries can be related to the symmetries and first integrals of SDEs. 
Later we show how these symmetries can be related to the symmetries of  the Kolmogorov  forward equation.

For SDEs~(\ref{sys}) the associated KBE  has the form 
\begin{equation}   \label{BKE}
- { \partial u \over \partial t }
=    f_i  (t,{\bf x}   )     { \partial  u  \over \partial x_i }       
+ {1\over 2}    g_{i \alpha} (t,{\bf x} ) g_{j \alpha} (t,{\bf x} )     { \partial ^2  u \over \partial x_i  \partial x_j }   .
\end{equation}
For symmetry analysis we rewrite it as 
\begin{equation}  \label{BK}
u_t
+ A_{ij} u _{x_i x_j}
+ B_k    u _{x_k} = 0 ,
\end{equation}
where
\begin{equation*}  
A_{ij} =  {1 \over 2} g_{ i \alpha }  g_{ j \alpha } ,
\qquad
B_k = f_k  .
\end{equation*}     
In what follows we will assume that $A_{ij}$ are not all zero.


\subsection{Determining equations}

Let us find Lie point symmetries
\begin{equation}   \label{symmetryBK}
X _{KB}    = \tau (t,{\bf x}, u) {\ddt}
+ \xi _i ( t,{\bf x}, u ) {\partial \over \partial {x_i}}
+ \eta ( t,{\bf x}, u ) {\ddu}
\end{equation}
which are admitted by the KBE.    
For our purpose we need the prolonged symmetry vector field 
\begin{equation}  
\mbox{\bf pr} ^{(2)}  X _{KB}    = \tau {\ddt}
+ \xi _i {\partial \over \partial {x_i}}
+ \eta {\ddu}  
+ \zeta _t {\partial \over \partial u_t } 
+ \zeta _i {\partial \over \partial u_{x_i} } 
+ \zeta _{ij} {\partial \over \partial u_{x_i x_j} }  , 
\end{equation}
where the coefficients are computed according to the standard prolongation formulas 
$$
\zeta_t  = D_t ( \eta ) - u_t  D_t  (  \tau )   - u_{x_j}  D_t  (  \xi_j  )  , 
$$
$$
\zeta_i  = D_i ( \eta ) - u_t  D_i  (  \tau )   - u_{x_j}  D_i  (  \xi_j  )  , 
$$
$$
\zeta_{ij}  = D_i ( \zeta_j  ) - u_{t x_j}   D_i  (  \tau )   - u_{x_j x_i}  D_i  (  \xi_j  )   . 
$$
Here $D_t$  and $D_i$ are total differentiation operators with respect to $t$ na $x_i$.

Infinitesimal invariance criteria~\cite{Ovs, Ibr1, Olver}
states that the application of the second prolongation of the operator $X _{KB}  $
to the second order PDE~(\ref{BK}) should be zero on the solutions of this PDE:
\begin{equation}   \label{criteria}
\left.
\mbox{\bf pr} ^{(2)}  X   _{KB}  ( u_t  + A_{ij} u _{x_i x_j}  + B_k    u _{x_k}    )
\right| _{(\ref{BK})}
= 0 .
\end{equation}



We review briefly  the derivation of the determining equations
for symmetries of the KBE. 
It is convenient to use notations
\begin{equation*}  
F_{,t} = { \partial F \over \partial t},
\qquad
F_{,u} = { \partial F \over \partial u}
\qquad
\mbox{and}
\qquad
F_{,i} = { \partial F \over \partial x_i}
\end{equation*}  
(the last notation uses indexes different from $t$ and $u$).

Equation~(\ref{criteria}) splits
for different spatial derivatives of $u$.
We obtain
\begin{equation*}  
 \tau _{,u}  ( A_{ij} A_{pq}   u _{ x_p x_q x_j} u _{ x_i}  )   = 0 
\end{equation*}  
for products of third derivatives with first derivatives that leads to 
\begin{equation*}  
 \tau _{, u }= 0  
\end{equation*}  
and
\begin{equation*} 
( A_{ij} \tau_{,i} )  ( A_{pq}   u _{ x_p x_q x_j}  )  = 0
\end{equation*}
for third derivatives that gives 
\begin{equation} \label{BK1}
A_{ij} \tau_{,i} = 0    . 
\end{equation}
Then, for products of second derivatives with first derivatives we get the equations
\begin{equation*}  
    \xi_{k,u}   ( A_{ij}   u _{ x_k x_i}  u _{x_j}   ) = 0  ,  
\end{equation*} 
which give 
\begin{equation*}  
    \xi_{i, u }   =    0  , 
\end{equation*}  
and  as coefficients for the second derivatives we obtain  equations 
\begin{equation} \label{BK5}
  \tau A_{ij,t}  + \xi_k A_{ij,k}
+ A_{ij} \left(  \tau_{,t}  + B_p  \tau _{,p } +  A_{pq} \tau _{,pq}  \right)
- A_{ik} \xi_{j,k }
- A_{kj} \xi_{i,k}
= 0   . 
\end{equation}

For products of first derivatives we obtain 
\begin{equation*}  
 \eta_{,uu} (   A_{ij}   u _{x_i}  u _{x_j}   ) = 0 .
\end{equation*}  
Therefore, 
\begin{equation*}  
\eta = \varphi  (t,{\bf x}) u + \psi  ( t, {\bf x})   . 
\end{equation*}  
Substituting it  into the rest of Eq.~(\ref{criteria}), we get 
\begin{multline} \label{BK6}
  \xi _{i,t} +  B_p  \xi _{i,p} +   A_{pq} \xi_{i,pq}
- \tau B_{i,t}  - \xi_p B_{i,p}
- 2 A_{ij}  \varphi  _{,j} 
\\
- B_{i} \left(  \tau _{,t} +  B_p   \tau_{,p} + A_{pq} \tau _{,pq}  \right)
= 0 , 
\end{multline}
\begin{equation} \label{BK7}
\ \varphi _{,t} +  B_k   \varphi _{,k} +  A_{ij}  \varphi  _{,ij}
= 0
\end{equation}
and
\begin{equation}     \label{BK9}
\psi _{,t}
+ B_k  \psi _{,k}  
+ A_{ij} \psi _{,ij}
 = 0  
\end{equation}  
for the terms with the first derivatives, the terms with $u$ and the rest, respectively.


We can summarize the obtained results using the operators   $D_0$   and $ D_{\alpha}$, 
which were given in~(\ref{intcond1_ito}).

\begin{theorem}   \label{result1_BK}
Lie point symmetries of KBE~(\ref{BKE}) are given by
\begin{enumerate}
\item
vector fields of the form
\begin{equation}    \label{symmetry_BK}
X _{KB} = \tau (t,{\bf x}) {\ddt}
+ \xi _i (t,{\bf x}) {\partial \over \partial {x_i}}
+ \varphi ( t,{\bf x} ) u    {\ddu}
\end{equation}
with coefficients satisfying equations 
\begin{subequations}       \label{theorem_conditions_0} 
\begin{gather} 
g _{ i \alpha   } D_{\alpha } (\tau ) = 0  ,      \label{theorem_conditions_1} 
\\
\begin{array}{l}
{ \displaystyle 
g_{i \alpha}
\left( D_{\alpha} ( \xi_j ) - X ( g_{ j \alpha} )
- {1 \over 2}  g_{ j \alpha} D_0 ( \tau ) \right)      }
\\
{ \displaystyle     \qquad
+  g_{j \alpha}
\left( D_{\alpha} ( \xi_i ) - X ( g_{ i \alpha} )
- {1 \over 2}  g_{ i \alpha} D_0 ( \tau ) \right)
= 0   } ,    
\end{array}  \label{theorem_conditions_2} 
\\
D_0 ( \xi_i ) - X ( f_i ) - f_i D_0 ( \tau )   =   g _{ i \alpha   } D_{\alpha } ( \varphi  )    ,    \label{theorem_conditions_3} 
\\
D_0 (  \varphi  )  = 0        \label{theorem_conditions_4} 
\end{gather}
\end{subequations}

and

\item
trivial symmetries
\begin{equation}    \label{symA1_BK}
X _{KB}     ^* = \psi   ( t,{\bf x} ) {\ddu} ,
\end{equation}
where the coefficient is an arbitrary solution of the KBE,
corresponding to the linear superposition principle.
\end{enumerate}
\end{theorem}

The proof follows from the previous discussion of the equations for symmetry coefficients. 
In particular, equations~(\ref{theorem_conditions_0}) for coefficients of the symmetry~(\ref{symmetry_BK}) 
represent equations  (\ref{BK1}), (\ref{BK5}), (\ref{BK6}) and (\ref{BK7}), 
which are rewritten with the help of the operators  $D_0$   and $ D_{\alpha}$. 
The coefficient of the symmetry~(\ref{symA1_BK}) satisfies equation~(\ref{BK9}).

\begin{remark}
We see that the determining equations~(\ref{theorem_conditions_0})   always have a particular   solution
\begin{equation*}  
\tau = 0  , \qquad    \xi _1 = ...  = \xi _n  = 0, \qquad      \varphi  = \mbox{const} . 
\end{equation*}  
It provides us with symmetry
\begin{equation}   \label{symA2}
X_0 = u{\ddu},
\end{equation}
corresponding to linearity of the KBE. 
\end{remark}


\subsection{Symmetries  of KBE  and symmetries of SDEs}

Now we can relate symmetries of the SDEs 
to the symmetries of the associated KBE.

\begin{theorem}   \label{result2}
Let operator $X$ of the form~(\ref{symmetry})
be a symmetry of the SDEs~(\ref{sys}),
then $X$ is also a symmetry of the  associated KBE. 
\end{theorem}

\noindent {\it Proof.}
From the determining equations~(\ref{cond2new})-(\ref{cond3new})
it follows that Eqs.~(\ref{theorem_conditions_1}) and Eqs.~(\ref{theorem_conditions_2})  hold. 
Choosing $  \varphi   \equiv 0 $,
which is always a solution of Eqs.~(\ref{theorem_conditions_3}) (if Eqs.~(\ref{cond1new}) hold) 
and~(\ref{theorem_conditions_4}),
we get $X$ as a symmetry of the KBE.
\hfill $\Box$

\medskip

We can also relate some symmetries of the KBE to first integrals of the SDEs.

\begin{theorem}    \label{result3}
Let SDEs~(\ref{sys}) possess a first integral $I(t, {\bf x})$,
then the  associated KBE    admits symmetry
\begin{equation}    \label{secondsymmetry}
Y = I(t, {\bf x}) u  {\ddu} .
\end{equation}
\end{theorem}

\noindent {\it Proof.}
It follows from Eqs.~(\ref{intcond1_I})
that the determining equations~(\ref{theorem_conditions_0})  for symmetries of the KBE  are satisfied.
\hfill $\Box$

\medskip

It is possible to state the converse results.

\begin{theorem}   \label{result4}
If KBE~(\ref{BKE}), which corresponds to SDEs~(\ref{sys}),
admits a symmetry ${X}$ of the form~(\ref{symmetry})
with coefficients satisfying equations~(\ref{cond2new}) and~(\ref{cond3new}),
then the symmetry $X$ is admitted by the SDEs.
\end{theorem}

\begin{theorem}     \label{result5}
If KBE~(\ref{BKE}), which correspond to SDEs~(\ref{sys}),
admits a symmetry of the form~(\ref{secondsymmetry})
and function $I(t, {\bf x})$ satisfies the equations~(\ref{intcond1_I_b}),
then $I(t, {\bf x})$ is a first integral of the SDEs.
\end{theorem}

The additional requirements of Theorems~\ref{result4} and~\ref{result5}  
are not surprising. 
They  specify the particular SDEs: 
the same KBE  can correspond to different SDEs,
which have the same drift coefficients $f_i$ and 
diffusion matrix 
$A_{ij} = {1\over 2} g_{i \alpha} g_{j \alpha} $.

\medskip

Finally, we summarize the results of this point 
by presenting four types of
Lie point symmetries of the Kolmogorov backward equation.
They are

\begin{enumerate}

\item

symmetries~(\ref{symA1_BK}) and~(\ref{symA2}) corresponding to linearity of the KBE 

\item

symmetries~(\ref{symmetry})  which are related to the symmetries of the SDEs

\item

symmetries~(\ref{secondsymmetry})    which are related to the first integrals of the SDEs

\item

the other symmetries, which are not related to the SDEs

\end{enumerate}


\subsection{Symmetries  of KBE and symmetries of KFE}

For SDEs~(\ref{sys}) the corresponding Kolmogorov  forward equation, 
which is also called Fokker--Planck equation~\cite{Risken},    
takes the form

\begin{equation}      \label{FP} 
{ \partial u \over \partial t }
= -  { \partial  \over \partial x_i }  ( f_i (t,{\bf x}) u )
+ {1\over 2}  { \partial ^2  \over \partial x_i  \partial x_j }
( g_{i \alpha} (t,{\bf x} ) g_{j \alpha} (t,{\bf x} ) u )   .
\end{equation}

The relation of symmetries of KFE  and the symmetries of the underlying SDEs 
was considered in several papers~\cite{Gae1, Unal1, Kozlov5}. 
The most general results were established in~\cite{Kozlov5}.  
They were based on the following description of the symmetries of the KFE.

\begin{theorem}   \label{result1_FK}
Lie point symmetries of KFE~(\ref{FP}) are given by
\begin{enumerate}
\item
vector fields of the form
\begin{equation}   \label{symmetry_FK}
X_{KF}  = \tau (t,{\bf x}) {\ddt}
+ \xi _i (t,{\bf x}) {\partial \over \partial {x_i}}
+ \chi   ( t,{\bf x} ) u    {\ddu}
\end{equation}
with coefficients satisfying equations

\begin{subequations} 
\begin{gather} 
g _{ i \alpha   } D_{\alpha } (\tau ) = 0  ,      
\label{FP_theorem_conditions_1}  
\\
\begin{array}{l}
{ \displaystyle   
g_{i \alpha}
\left( D_{\alpha} ( \xi_j ) - X ( g_{ j \alpha} )
- {1 \over 2}  g_{ j \alpha} D_0 ( \tau ) \right)      }
\\
{ \displaystyle     \qquad 
+  g_{j \alpha}
\left( D_{\alpha} ( \xi_i ) - X ( g_{ i \alpha} )
- {1 \over 2}  g_{ i \alpha} D_0 ( \tau ) \right)
= 0   } ,    
\end{array}  \label{FP_theorem_conditions_2} 
\\
\label{FP9}
Q = \chi     + \xi_{i,i} + \tau_{,t} - D_0 ( \tau )  , 
\end{gather}
\end{subequations}
where function $Q (t,{\bf x}) $ is a solution of equations 
\begin{subequations} 
\begin{gather}
D_0 ( \xi_i ) - X ( f_i ) - f_i D_0 ( \tau )   = -   g _{ i \alpha   } D_{\alpha } ( Q  )    ,    
\label{FP_theorem_conditions_3} 
\\ 
D_0 (  Q  )  = 0   ,      
\label{FP_theorem_conditions_4} 
\end{gather}
\end{subequations}

and

\item
trivial symmetries
\begin{equation}      \label{symA1_FK}
X_{KF}    ^* = \psi ( t,{\bf x} ) {\ddu} ,
\end{equation}
where the coefficient is an arbitrary solution of the KFE,   
corresponding to the linear superposition principle.
\end{enumerate}
\end{theorem}

By direct comparison of the determining equations 
given in Theorems~\ref{result1_BK}  and~\ref{result1_FK} 
we can establish the following result.

\begin{theorem}         \label{symmetry_relation} 
Let us consider KBE~(\ref{BKE}) and KFE~(\ref{FP}) corresponding to the same SDEs~(\ref{sys}).  
The KBE   admits symmetry~(\ref{symmetry_BK})  
if and only if  
the KFE admits symmetry~(\ref{symmetry_FK})  
and 
\begin{equation*} 
\varphi    +   \chi     =   D_0 ( \tau )    -   \tau_{,t}   -     \xi_{i,i}      . 
\end{equation*} 
\end{theorem}

\noindent {\it Proof.} 
The result follows from the observation that 
the sets of variables $  ( \tau , \xi _1, ..., \xi_  n  ,  \varphi )$   and   $  ( \tau , \xi _1, ... , \xi_  n ,   - Q   )$   
satisfy the same equations. 
\hfill $\Box$

\medskip

\begin{corollary} 
Let us consider KBE~(\ref{BKE}) and KFE~(\ref{FP}) corresponding to the same SDEs~(\ref{sys}).  
The KBE   admits symmetry~(\ref{secondsymmetry})  
if and only if  
the KFE admits the same symmetry. 
\end{corollary}


\section{Scalar SDEs and   $(1+1)$-dimensional  Kolmogorov equations.}

\label{section_scalar}

Let us illustrate how one can use symmetries of the scalar SDEs  
\begin{equation}       \label{example_SDE_0}
 dx = f(t,x)  dt  + g (t,x)  dW (t)     , \qquad      g (t,x)  { \not \equiv } 0 
\end{equation}  
to find symmetries of the Kolmogorov backward equation 
\begin{equation}    \label{example_BK_0}
 -  { \partial u \over \partial t }  
=      f(t,x)  { \partial  u \over \partial x }       
 + {1\over 2}    G (t,x)     { \partial ^2  u \over \partial x^2 }      , 
\qquad
  G  (t,x)  = g^2    (t,x) \geq 0 , {\not\equiv} 0      . 
\end{equation}

Lie point symmetries of the KBE~(\ref{example_BK_0}) are described by Theorem~\ref{result1_BK}. 
They  are   symmetries 
\begin{equation}
X _{KB} = \tau (t,{x}) {\ddt}
+ \xi (t,{ x}) {\partial \over \partial {x}}
+ \varphi ( t,{x} ) u    {\ddu}
\end{equation}
with coefficients satisfying equation 
\begin{subequations}      \label{example_condition_0}
\begin{gather}
 D_{W} (\tau ) = 0  ,        \label{example_condition_1}
 \\
D_{W} ( \xi ) - X ( g )
- {1 \over 2}  g  D_0 ( \tau ) 
= 0  ,     \label{example_condition_2}
\\
D_0 ( \xi ) - X ( f ) - f D_0 ( \tau )   =   g  D_{W} ( \varphi  )    ,    \label{example_condition_3}
\\
D_0 (  \varphi  )  = 0   ,      \label{example_condition_4}
\end{gather}
\end{subequations}
where 
\begin{equation} 
D_0 
=  { \partial   \over \partial t }
+  f   { \partial  \over \partial x }  
+ {1\over 2} g^2  { \partial ^2  \over \partial x ^2   }  ,
\qquad
D_{W}   = g       { \partial  \over \partial x }      ,      
\end{equation}
and trivial symmetries~(\ref{symA1_BK}). 
Note that from~(\ref{example_condition_1})   we get $ \tau  = \tau (t) $.

In the general case the KBE~(\ref{example_BK_0})  has only  symmetries related to its linearity, 
namely 
\begin{equation} \label{linearityKB}
X_{KB}  ^* = \psi (t,x) {\ddu}
\qquad
\mbox{and}
\qquad
X_0 = u {\ddu} ,
\end{equation}
where $\psi (t,x)$ is an arbitrary solution of the KBE. 
For particular cases   $ f(t,x) $  and $G(t,x) $ there can be additional symmetries. 


\subsection{Symmetries of KBE via symmetries of SDEs}

Lie group classification of the scalar SDE~(\ref{example_SDE_0})   
was carried out in~\cite{Kozlov1} by direct method. 
Alternatively, one can obtain this Lie group classification 
with the help of real Lie algebra realizations by vector fields. 
It was done  in~\cite{Kozlov2}.

In the general case the SDE~(\ref{example_SDE_0})  has no symmetries. 
Therefore, the KBE~(\ref{example_BK_0}) 
admits only symmetries~(\ref{linearityKB}) corresponding to its linearity. 
We shall go through the cases of the Lie group classification  of the scalar SDEs~(\ref{example_SDE_0})  
and find the symmetries of the corresponding KBEs. 
It should be noted that we can always choose a representative SDE for each equivalence symmetry class 
in the form 
\begin{equation}       \label{representative}
 dx = f(t,x)  dt  +  dW (t)    
\end{equation}  
because we can perform the variable change 
\begin{equation*}    
x \rightarrow     \int  {dx   \over g (t,x) }    . 
\end{equation*}
The corresponding  KBE is also simplified. It takes the  form 
\begin{equation}    \label{BK_0}
 -  { \partial u \over \partial t }  
=      f(t,x)  { \partial  u \over \partial x }       
 + {1\over 2}     { \partial ^2  u \over \partial x^2 }      . 
\end{equation}

\subsubsection{SDE with one symmetry}

The equivalence  class of the SDEs admitting only one symmetry  
\begin{equation}        \label{symmetry_SDE_1}
X_{1}= { \displaystyle {\ddt}  }    
\end{equation}  
can be represented by the equation  
\begin{equation}       \label{example_SDE_1}
   dx = f(x)  dt  +  dW   (t)      . 
\end{equation}

The corresponding KBE 
\begin{equation}       \label{example_BK_1}
- u _t    =        f(x) u _x     +     {1\over 2}   u  _{xx}   
\end{equation}  
admits symmetries~(\ref{linearityKB}) and~(\ref{symmetry_SDE_1}).

\subsubsection{SDE with two symmetries}

For the SDEs admitting two symmetries 
\begin{equation}     \label{symmetry_SDE_2}
X_{1}= { \displaystyle {\ddt}  }  ,
\qquad
X_{2}= { \displaystyle 2t {\ddt} +  x {\ddx}  }   
\end{equation}  
one can chose the following  representative equation 
\begin{equation}      \label{example_SDE_2}
dx = { \displaystyle { A \over x } } dt  +      dW  (t) , \qquad   A \neq 0     .
\end{equation}

For the KBE 
\begin{equation}       \label{example_BK_2}
 -  u _t   =   { A \over x } u _x     +   {1 \over 2}   u  _{xx} 
\end{equation}  
we get two {sub}cases:

\begin{enumerate} 

\item

$ A  \neq  1     $

In addition to symmetries~(\ref{linearityKB}) and~(\ref{symmetry_SDE_2}) 
the  KBE   admits    symmetry   
\begin{equation}      \label{symmetry_SDE_2a}
Y_1 =  2t^2  {\ddt} + 2tx{\ddx}  
+  \left(     {x^2 }  -  \left( 1 +  { 2 A  } \right)    t \right) u {\ddu}     . 
\end{equation}



\item 

$ A = 1 $ 

In this particular case  the KBE possesses 
symmetries~(\ref{linearityKB}), (\ref{symmetry_SDE_2}), (\ref{symmetry_SDE_2a}) and 
\begin{equation}      \label{symmetry_SDE_2b}
Y_2 =    t {\ddx}    +  \left( {x }  -  { t \over x } \right) u {\ddu} , 
\qquad
Y_3 =     {\ddx}     -    { u  \over x }  {\ddu}    . 
\end{equation}


\end{enumerate}


\subsubsection{SDE with three symmetries}

Scalar SDEs can admit at most three symmetries.  
The equivalence class for SDEs with three symmetries can be represented by the equation   
\begin{equation}       \label{example_SDE_3}
   dx =  dW   (t)    , 
\end{equation}  
which admits symmetries 
\begin{equation}      \label{symmetry_SDE_3}
X_{1}= { \displaystyle {\ddt}  }  ,
\qquad
X_{2}= { \displaystyle 2t {\ddt} +  x {\ddx}  } ,
\qquad
X_{3}= { \displaystyle {\ddx}   .  }
\end{equation}

In this case we get the KBE 
\begin{equation}        \label{example_BK_3}
    -  u _t  =    {1 \over 2}   u  _{xx}  
\end{equation}  
admits symmetries~(\ref{linearityKB}), (\ref{symmetry_SDE_3}) and  
\begin{equation}     \label{symmetry_SDE_3b}
Y_1 = t {\ddx}  +  { x u } {\ddu}   ,  
\qquad
Y_2 = 2t^2  {\ddt} + 2tx {\ddx}   +   ( {x^2 } -  t  ) u {\ddu}    . 
\end{equation}



\begin{remark} 
The KBE~(\ref{example_BK_2})  with $A = 1$, namely the equation 
\begin{equation*}  
-  u _t     =     { 1 \over x } u  _x   +  { 1 \over 2}   u  _{xx}    ,
\end{equation*}  
can be transformed into the KBE~(\ref{example_BK_3})  
by the change of the dependent variable
\begin{equation*} 
u =  { 1 \over x }  \bar{u} .
\end{equation*}  
However, the SDE~(\ref{example_SDE_2}) with  $A = 1$ 
cannot be transformed into the SDE~(\ref{example_SDE_3}). 
\end{remark}

We cannot expect that Lie group classification of SDEs will provide us with Lie group classification 
of the associated KBE. 
Indeed, it gives only partial results on the symmetries of specified KBEs as we will see in the next point.


\subsection{Lie group classification of  $(1+1)$-dimensional KBE}

Lie group classification of  the $(1+1)$-dimensional  KFE  
\begin{equation} \label{FP11}
{ \partial u \over \partial t }
= -  { \partial  \over \partial x }  ( f(t,x) u )
+ {1\over 2}  { \partial ^2  \over \partial x ^2   }
( G (t,x)  u )  
\end{equation} 
is knows~\cite{Spichak}. 
It can be used to obtain Lie group classification of  the $(1+1)$-dimensional  KBE   
with the help of Theorem~\ref{symmetry_relation}, 
which relates symmetries of the Kolmogorov backward and Kolmogorov forward equations.

In addition to the symmetries 
\begin{equation}      \label{linearityKF}
X_{KF} ^* = \psi (t,x) {\ddu}
\qquad
\mbox{and}
\qquad
X_0 = u {\ddu} ,
\end{equation}
where $\psi (t,x)$ is an arbitrary solution of the KFE, 
the   KBE~(\ref{FP11})      can admit   0,  1, 3  or  5 symmetries. 
Due to Theorem~\ref{symmetry_relation}  we get the same results for the KBE~(\ref{example_BK_0}). 
It can admit 0,  1, 3  or  5 symmetries  in addition to symmetries~(\ref{linearityKB}).

Lie group classification of the KBE (obtained with the help of Lie group classification of the KFE) 
can be compared with results of the previous point. We find out the following.

\begin{itemize}

\item

Using Lie group classification of the scalar SDEs,  we obtain correct description 
of the equivalence classes for KBEs admitting 0, 1  and 5 symmetries in addition to the symmetries~(\ref{linearityKB}). 
These equivalence classes  are represented by 
the equations~(\ref{BK_0}), (\ref{example_BK_1}) and~(\ref{example_BK_3}), respectively.  

\item

However, we do not obtain the correct description of the equivalence class for 
the  KBEs admitting 3 additional  symmetries. 
It is easy to see from the next  theorem.

\end{itemize}


\begin{theorem} (\cite{Spichak})
KFEs~(\ref{FP11}) admitting three symmetries     in addition to  the linearity symmetries~(\ref{linearityKF}) 
 can be transformed into the equation
\begin{equation}  \label{foureq}
u_t = ( 2 k'(x)u )_x + u_{xx} ,
\end{equation}
where $k(x)$ is a solution of the equation
\begin{equation}  \label{fourcond}
k'' - (k') ^2 = {\lambda \over x^2 } , \qquad \lambda \neq 0 .
\end{equation}
\end{theorem}

In addition to the symmetries~(\ref{linearityKF}) the equation~(\ref{foureq})
admits operators
\begin{equation*} 
X_1 =  {\ddt}    ,
\qquad
X_2 = 2t {\ddt} +  x {\ddx}   - x k'(x) u {\ddu}  ,
\end{equation*} 
\begin{equation*} 
X_3 = 4 t^2  {\ddt} + 4tx{\ddx}
- ( x^2 + 4t + 4 t x k'(x) ) u {\ddu} ,
\end{equation*} 
where $k(x)$ is a solution of Eq.~(\ref{fourcond}).

Using Theorem~\ref{symmetry_relation}, 
which relates symmetries of the Kolmogorov backward and Kolmogorov forward equations, 
we state a similar result for the KBE.

\begin{corollary}
KBEs~(\ref{example_BK_0}) admitting three  symmetries     in addition to  the linearity symmetries~(\ref{linearityKB}) 
can be transformed into the equation
\begin{equation}  \label{fourKB}
- u_t  =  -   2 k'(x)  u _x +   u_{xx}  ,
\end{equation}
where $k(x)$ is a solution of the equation~(\ref{fourcond}). 
\end{corollary}

In addition to the symmetries~(\ref{linearityKB}) the equation~(\ref{fourKB})
admits operators
\begin{equation*} 
X_1 =  {\ddt}    ,
\qquad
X_2 = 2t {\ddt} +  x {\ddx}   +  x k'(x) u {\ddu}  ,
\end{equation*} 
\begin{equation*} 
X_3 = 4 t^2  {\ddt} + 4tx{\ddx}
 +  ( x^2 + 4 t  x k'(x)   )    u {\ddu} ,
\end{equation*} 
where $k(x)$ is a solution of Eq.~(\ref{fourcond}).  
For
\begin{equation*} 
k'(x) \neq{ A  \over x}
\end{equation*} 
symmetry $X_2$ (up to factorization by $X_0$) is beyond the framework
based on the symmetries of the underlying SDE. 
Thus, with the help of the scalar SDE classification
we get only a {sub}case of the class of KBEs admitting three additional symmetries.

Therefore, using the Lie group classification of the scalar SDEs, 
we get {\it partial} results of the Lie group classification of the $(1+1)$-dimensional KBE. 
The same  was observed for the $(1+1)$-dimensional  KFE in~\cite{Kozlov6}.


\subsection{The  geometric Brownian motion equation}

Let us examine the geometric Brownian motion~\cite{Oks} 
\begin{equation}       \label{geometric}
d x = \alpha x d t + \sigma x d W (t)    , 
\qquad   
\sigma \neq 0 
\end{equation}    
as a theory application. 
This SDE is an important model for stochastic prices in economics.

The SDE~(\ref{geometric}) admits three symmetries of the form~(\ref{symmetry}), namely  
\begin{equation}        \label{symmetries _geometrical} 
X_1 
=  {\ddt}  , 
\qquad 
X_2   
=  2t {\ddt}  
+ \left(    \left(  \alpha - { \sigma^2  \over 2 } \right) t x   + x \ln x   \right)    {\ddx}  , 
\qquad
X_3 
=  x {\ddx}
\end{equation}    
and has no  first integrals.

The associated   
Kolmogorov backwards equation~(\ref{example_BK_0}) takes the form 
\begin{equation}        \label{KB_geometric} 
 -  { \partial u \over \partial t }  
=       \alpha x   { \partial  u \over \partial x }       
 + {1\over 2}   \sigma^2 x^2  { \partial ^2  u \over \partial x^2 }      . 
\end{equation}
It admits the trivial symmetries~(\ref{linearityKB}) because of its linearity. 
Theorem~\ref{result2} states that the symmetries~(\ref{symmetries _geometrical}) 
are also admitted by this  KBE.  
Direct computation of the symmetries of the form~(\ref{transformations}) provides us 
with the two additional symmetries 
\begin{subequations}    
\begin{gather}
Y_1    
=  t x {\ddx}  
-  { 1 \over \sigma ^2 }   \left(    \left(  \alpha - { \sigma^2  \over 2 } \right) t  -  \ln x  \right)  u  {\ddu}  , 
\\
Y_2    
=  2 t^2 {\ddt}  
+   2 t x \ln  x {\ddx}  
+    \left(     { 1 \over \sigma ^2 }   \left(   \left(  \alpha - { \sigma^2  \over 2 } \right) t  - \ln x   \right)^2   - t       \right)   
  u {\ddu} . 
\end{gather}
\end{subequations}


The Kolmogorov forward  equation~(\ref{FP11}) for the geometric Brownian motion equation 
takes the form 
\begin{equation}       \label{KF_geometric} 
   { \partial u \over \partial t }  
=   -    \alpha   { \partial  \over \partial x }   ( x u )   
 + {1\over 2}   \sigma^2   { \partial ^2   \over \partial x^2 }  ( x^2 u )        . 
\end{equation}
It is invariant with respect to symmetries~(\ref{linearityKF}). 
The other symmetries can be obtained with the help of Theorem~\ref{symmetry_relation}   
and the symmetries of the   Kolmogorov backward equation~(\ref{KB_geometric}).  
We find  
$$
X_1 
=  {\ddt} , 
\qquad 
X_2   
=  2t {\ddt}  
+ \left(    \left(  \alpha - { \sigma^2  \over 2 } \right) t x  +  x \ln x    \right)    {\ddx} 
-   \left(   \left(  \alpha - { \sigma^2  \over 2 } \right) t  + \ln x    \right)    u {\ddu}   , 
$$
$$
X_3 
=  x {\ddx}  , 
\qquad 
Y_1    
=  t x {\ddx}  
+  \left(  { 1 \over \sigma ^2 }   \left(     \left(  \alpha -  { \sigma^2  \over 2 } \right) t   -  \ln x    \right) 
- t \right)   u  {\ddu}   , 
$$
$$
Y_2   
=  2 t^2 {\ddt}  
+   2 t x \ln  x {\ddx}  
-    \left(     { 1 \over \sigma ^2 }   \left(    \left(  \alpha - { \sigma^2  \over 2 } \right) t - \ln x \right)^2   
+  2 t  \ln x + t       \right)  u   {\ddu}   .
$$

\begin{remark}
Let us note that symmetries of SDEs can be used to find symmetries of the associated     
Kolmogorov forward  equation~\cite{Kozlov5}. 
A symmetry~(\ref{symmetry}) of the SDEs~(\ref{sys})  provides with the symmetry 
\begin{equation}   \label{firstsymmetry}
\bar{X} = X  + ( D_0 ( \tau )  -   \tau_{,t} - \xi_{i,i}   )  u  {\ddu}  
\end{equation}
admitted by the associated KFE. 
This results can be used to find the symmetries  $X_1$,  $X_2$ and  $X_3$ 
for the   KFE~(\ref{KF_geometric}).  
\end{remark}


\end{document}